\documentstyle[12pt]{article}

\begin{document}

\title{Ward Identities, $B\rightarrow \rho $ Form Factors and $\left| V_{ub}\right| 
$}
\author{Amjad Hussain Shah Gilani \thanks{%
E-mail address: ahgilani@yahoo.com} \\
Department of Physics and National Center for Physics,\\
Quaid-e-Azam University, Islamabad 45320, Pakistan. \and Riazuddin \thanks{%
E-mail address: ncp@comsats.net.pk} \\
Department of Physics, \\
King Fahd University of Petroleum and Minerals,\\
Dhahran 31261, Saudi Arabia \\
and \\
National Center for Physics, Quaid-e-Azam University, \\
Islamabad 45320, Pakistan. \and T.A. Al-Aithan \thanks{%
E-mail address: alaithan@kfupm.edu.sa} \\
Department of Physics, \\
King Fahd University of Petroleum and Minerals,\\
Dhahran 31261, Saudi Arabia.}
\date{}
\maketitle

\begin{abstract}
The exclusive FCNC beauty semileptonic decay $B\rightarrow \rho $ is studied
using Ward identities in a general vector meson dominance framework,
predicting vector meson couplings involved. The long distance contributions
are discussed which results to obtain form factors and $\left| V_{ub}\right| 
$. A detailed comparison is given with other approaches.
\end{abstract}

\newpage\ 

\section{Introduction}

The semileptonic and radiative decays of heavy mesons play an outstanding
role for determination of the parameters of standard model, in particular
the quark mixing parameters.

The upcomming and currently operating $B$ factories BaBar at SLAC, Belle at
KEK, LHCB at CERN and B-TEV at Fermilab as well as the planned $\tau $-Charm
factory CLEO at Cornell make precision tests of Standard Model (SM) and
beyound SM ever more promising \cite{expt}. Especially, a strigent test on
the unitarity of the Cabibo-Kobayashi-Maskawa (CKM) mixing matrix \cite{CKM}
in the SM will be made by these facilities. Accurate analysis of exclusive
semileptonic $B$ decays are thus strongly demanded for such precision tests.
Beauty decays proceeding through the flavour changing neutral current (FCNC)
transition have been identified as a valuable source of information on
standard model and its extensions \cite{Shifman} (for a review and complete
set of refrences see \cite{PRs245-259,RMP67-893,RMP68-1125}). One of the
physics programs at the $B$ factories is the exclusive $B$ decays induced by
FCNC transitions. These decays are forbidden at the tree level in the SM and
occurs at the lowest order only through one-loop (penguin) diagrams. The
experimental tests of exclusive decays are much easier than those of
inclusive ones.

The theoretical understanding of weak decays of hadrons, and the
measurements of the corresponding CKM matrix elements are consistently
hampered by the presence of long distance QCD effects that are responsible
for the binding of quarks into hadrons. These effects are hard to evaluate
in a model independent way, and so tend to bring large uncertainities to the
theoretical predictions for the weak decay amplitudes. They appear in the
calculation of the matrix elements of the weak Hamiltonian operators,
between the initial and final hadronic states.

The dynamical content of hadronic current matrix elements is described by
Lorentz-invariant form factors. The calculation of these form factors
requires a non-perturbative treatment and are source of large theoretical
uncertainities. Knowledge of these form factors is essential for the
description of semileptonic and non-leptonic weak decay processes and in
particular for the experimental determination of CKM matrix elements \cite
{CKM}. Many theoretical tools have been applied for the determination of
form factors, i.e. quark models, QCD sum rule, and lattice calculations. The
machinery of sum rules, lattice calculations has the advantage to be
directly based on the QCD Lagrangian, but is quite involved. Quark models,
on the other hand are less directly connected with the QCD Lagrangian, but
give a vivid picture of what is going on and allow an easy application to
different processes and at quite different kinematical regions. However,
models presented so far lacked full relativistic covariance with respect to
quark spins, and it is the spin structure which determines the ratio of
different form factors we are concerned. The knowledge of these ratios is of
paramount importance for the analysis of experimentally measured decays. So,
the relations between different form factors that will hold under certain
conditions or approximations can then be very useful. One of the motivations
of the present work is to make use of Ward Identities to relate various form
factors in a model independent way. Such relations will reduce the number of
uncertain quantities, and improve the accuracy of the theoretical
predictions. Moreover, they may help us understand better the general
features of the underlying long distance QCD effects.

As mentioned earlier, the theoretical understanding of exclusive decays is
complicated mainly due to the non-perturbative hadronic form factors entered
in the long distance nonperturbative contributions. These calculations of
hadronic form factors for semileptonic $B$ decays have been investigated by
various theoretical approches. Form factors based on lattice calculations 
\cite{PLB416-392,lattice}, and on light cone sum rules \cite
{PRD55-5561,LCSR,PRD58-094016}, currently have uncertainties in the 15\% to
20\% range. A variety of quark model calculations \cite
{ZPC29-637,ISGW2,PLB436-344,quarkmodel,ZPC38-511,amjadgilani} exist. A
number of other approaches \cite{other,HQS1992}, such as dispersive bounds
and experimentally constrained models based on Heavy Quark Symmetry (HQS)
also exists. In the standard model, the short distance contributions are
dominated by the top-quark and long distance contributions by form factors.

The aim of the present work, as mentioned earlier, is to relate the various
form factors in a model independent way through Ward identities. This
enables us to make a clean separation between non--pole and pole type
contributions, the $q^2\rightarrow 0$ behaviour of the former is known in
terms of a universal function $\xi _{\perp }\left( q^2\right) \equiv
g_{+}\left( q^2\right) $ introduced in the large energy effective theory
(LEET) of heavy $\left( B\right) $ to light $\left( \rho \right) $ form
factors \cite{PRD60-014001}. The corrections of order $(m/M)^2$, where $m$
is the light meson mass and $M$ is the heavy meson mass, in LEET were
obtained in \cite{PRD64-094022}. The LEET approach was first proposed by
Dugan and Grinstein \cite{PLB255-583} and further developed by Aglietti \cite
{PLB292-424}.

The residue of the pole is then determined in a self consistent way in terms
of $g_{+}\left( 0\right) $ or $\xi _{\perp }\left( 0\right) $ [a sort of
duality]. Thus our approach has a predictive power as the residue of the
pole will give information about the couplings of $B^{*}\left( 1^{-}\right) $
and $B_A^{*}\left( 1^{+}\right) $ with $B\rho $ channel. The form factors
are then determined in terms of $g_{+}\left( 0\right) $, $A_0\left( 0\right) 
$ (for which much information is already known) and pole masses $M_{B^{*}}$, 
$M_{B_A^{*}}$ (where $M_{B^{*}}$ is known and estimates exists for $%
M_{B_A^{*}}$), and $M^{\prime }/M$ [where $M^{\prime }$ is the radial
excitation of $M$ ($B^{*}$ and $B_A^{*}$)]. This ratio is 1.14 (1.12) \cite
{PRD21-203}.

The information about the hadronic form factors thus obtained are used to
calculate the decay rates and the CKM matrix element $V_{ub}$. Because $%
\left| V_{ub}\right| $, the smallest element in the CKM mixing matrix, plays
a crucial role in the examination of the unitarity constraints and the
fundamental questions on which the constraints can bear \cite{pdg1}. To
extract $\left| V_{ub}\right| $ from exclusive channel, the form factors for
the channel must be known. The form factor normalization dominates the
uncertainty on $\left| V_{ub}\right| $. The $q^2$ dependence of the form
factors, which is needed to determine the experimental efficiency, also
contribute to the uncertainty, but at much reduced level. There have been
two exclusive $\left| V_{ub}\right| $ analyses by the CLEO Collaboration: a
simulatneous measurement of the $B\rightarrow \pi l\bar{\nu}$ and the $%
B\rightarrow \rho l\bar{\nu}$ transitions \cite{CLEO1}, and second
measurement of the $B\rightarrow \rho l\bar{\nu}$ rate \cite{PRD61-052001}.
The branching fractions obtained were 
\begin{eqnarray*}
{\cal B}\left( B^0\rightarrow \rho ^{-}l^{+}\nu \right) &=&\left( 2.5\pm
0.4_{-0.7}^{+0.5}\pm 0.5\right) \times 10^{-4}\,\,{\cite{CLEO1}} \\
{\cal B}\left( B^0\rightarrow \rho ^{-}l^{+}\nu \right) &=&\left( 2.69\pm
0.41_{-0.40}^{+0.35}\pm 0.5\right) \times 10^{-4}\,\,{\cite
{PRD61-052001}}.
\end{eqnarray*}
The results of the two analyses are largely statistically independent, and
they have been combined 
\[
\left| V_{ub}\right| =\left( 3.25\pm 0.14_{-0.29}^{+0.21}\pm 0.55\right)
\times 10^{-3}, 
\]
where the errors arise from statistical, experimental systematic, and form
factor uncertainities, respectively. A recent study by BaBar \cite{ex0301001}%
, obtained the following results for the branching ratios and $\left|
V_{ub}\right| $ and 
\begin{eqnarray*}
{\cal B}\left( B^0\rightarrow \rho ^{-}l^{+}\nu \right) &=&\left( 3.29\pm
0.42\pm 0.47\pm 0.60\right) \times 10^{-4}{,}\, \\
\left| V_{ub}\right| &=&\left( 3.64\pm 0.22\pm 0.25_{-0.56}^{+0.39}\right)
\times 10^{-3},
\end{eqnarray*}
where the uncertainities are statistical, systematic, and theoretical
respectively; and CLEO \cite{ex0304019}, 
\begin{eqnarray*}
{\cal B}\left( B^0\rightarrow \rho ^{-}l^{+}\nu \right) &=&\left( 2.17\pm
0.34_{-0.54}^{+0.47}\pm 0.41\pm 0.01\right) \times 10^{-4}{,} \\
\left| V_{ub}\right| &=&\left( 3.00\pm
0.21_{-0.35}^{+0.29}\,_{-0.38}^{+0.49}\pm 0.28\right) \times 10^{-3},
\end{eqnarray*}
where the errors in CLEO\ analyses are statistical, experimental systematic,
theoretical systematicbased on lattice QCD and LCSR uncertainities, and $%
\rho l\nu $ form factor shape uncertainity, respectively.

The future for the exclusive determinations of $\left| V_{ub}\right| $
appears promising. For both lattice and B factories, $B\rightarrow \pi l\nu $
appears to be a golden mode for future precise determination of $\left|
V_{ub}\right| $ while the $B\rightarrow \rho l\nu $ mode will be more
problematic for high precision because the broad width introduces both
experimental and theoretical difficulties.

This paper is organized as follows: In section \ref{cmes}, we give the $%
B\left( p\right) \rightarrow V\left( k,\varepsilon \right) l\bar{\nu}$
current matrix elements. In section \ref{WIs}, we discuss the Ward
Identities and develop the relations between form factors which result in to
reduce the number of unknown quantities. In section \ref{pcanddff}, pole
contribution of various form factors are discussed and relations among
different decay constants with the help of Ward Identities are obtained. In
section \ref{ffandcc}, form factors are obtained which are necesary for the
calculation of decay width and branching ratio. In section \ref
{sec-decay-distribution}, we discuss the decay distribution in terms of
helicity amplitudes and the extraction of $\left| V_{ub}\right| $ from the
experimental data. In section \ref{numerical-analyses}, numerical analyses
is made and our predictions regarding various observables are compared with
other approaches. Section \ref{conclusion}, summarizes our conclusions and
predictions.

\section{Current Matrix Elements \label{cmes}}

We are interested in the exclusive semileptonic decays of pseudoscalar meson
($B$) into vector meson ($\rho $). The structure of hadronic current in
semileptonic decay must be constructed from the available four-vectors,
which are momenta and spin-polarization vectors. The Lorentz-vector or axial
vector quantities thus formed have Lorentz invariant coefficients (form
factors) that are functions of $q^2=\left( p-k\right) ^2$. The polarization
vector of the meson $\rho \left( k,\varepsilon \right) $ leads to a hadronic
current with three form factors (in the limit of zero charged lepton mass).

In the case of $B\left( p\right) \rightarrow \rho \left( k,\varepsilon
\right) l\bar{\nu}$ decay, there are seven form factors for pseudoscalar to
vector transitions, i.e. $V$, $A_{0,1,2}$, and $F_{1,2,3}$ \cite
{PRs245-259,RMP67-893,ZPC29-637} 
\begin{eqnarray}
\left\langle V\left( k,\varepsilon \right) \left| V^\mu \right| B\left(
p\right) \right\rangle &=&\frac{2i\epsilon ^{\mu \nu \alpha \beta }}{M_B+M_V}%
\varepsilon _\nu ^{*}k_\alpha p_\beta V\left( q^2\right)  \label{hcV} \\
\left\langle V\left( k,\varepsilon \right) \left| A^\mu \right| B\left(
p\right) \right\rangle &=&\left( M_B+M_V\right) \varepsilon ^{*\mu
}A_1\left( q^2\right)  \nonumber \\
&&-\frac{\varepsilon ^{*}\cdot q}{M_B+M_V}\left( k+p\right) ^\mu A_2\left(
q^2\right)  \nonumber \\
&&-2M_V\frac{\varepsilon ^{*}\cdot q}{q^2}q^\mu \left[ A_3\left( q^2\right)
-A_0\left( q^2\right) \right]  \label{hc1}
\end{eqnarray}
where $V^\mu =\bar{u}\gamma ^\mu b$, $A^\mu =\bar{u}\gamma ^\mu \gamma ^5b$,
and 
\begin{equation}
A_3\left( q^2\right) =\frac{M_B+M_V}{2M_V}A_1\left( q^2\right) -\frac{M_B-M_V%
}{2M_V}A_2\left( q^2\right)  \label{A3qsq}
\end{equation}
with $A_3\left( 0\right) =A_0\left( 0\right) $, 
\begin{eqnarray}
\left\langle V\left( k,\varepsilon \right) \left| \bar{u}i\sigma ^{\mu \nu
}q_\nu b\right| B\left( p\right) \right\rangle &=&-i\epsilon ^{\mu \nu
\alpha \beta }\varepsilon _\nu ^{*}k_\alpha p_\beta F_1\left( q^2\right)
\label{hc2} \\
\left\langle V\left( k,\varepsilon \right) \left| \bar{u}i\sigma ^{\mu \nu
}\gamma ^5q_\nu b\right| B\left( p\right) \right\rangle &=&\left[ \left(
M_B^2-M_V^2\right) \varepsilon ^{*\mu }-\left( \varepsilon ^{*}\cdot
q\right) \left( k+p\right) ^\mu \right] F_2\left( q^2\right)  \nonumber \\
&&+\left( \varepsilon ^{*}\cdot q\right) \left[ q^\mu -\frac{q^2}{M_B^2-M_V^2%
}\left( k+p\right) ^\mu \right] F_3\left( q^2\right)  \label{hc3}
\end{eqnarray}
with $F_1\left( 0\right) =2F_2\left( 0\right) $.

Again the term proportional to $q^\mu $ in Eqs. (\ref{hc1}--\ref{hc3}) only
plays an important role for the case $l=\tau $ and is not relevent for the
rate in case of produced lepton pair $l=\mu $, $e$. The same is true for the
form factor $F_3$. However, it does play a considerable role in the Ward
identity.

\section{Ward Identities\label{WIs}}

The Ward Identities, which are derived in appendix A, relate various form
factors \cite{ZPC48-55} 
\begin{eqnarray}
\left\langle V\left( k,\varepsilon \right) \left| \bar{u}i\sigma ^{\mu \nu
}q_\nu b\right| B\left( p\right) \right\rangle &=&-\left( m_b+m_q\right)
\left\langle V\left( k,\varepsilon \right) \left| \bar{u}\gamma ^\mu
b\right| B\left( p\right) \right\rangle  \label{WI1a} \\
\left\langle V\left( k,\varepsilon \right) \left| \bar{u}i\sigma ^{\mu \nu
}\gamma ^5q_\nu b\right| B\left( p\right) \right\rangle &=&\left(
m_b-m_q\right) \left\langle V\left( k,\varepsilon \right) \left| \bar{u}%
\gamma ^\mu \gamma ^5b\right| B\left( p\right) \right\rangle  \nonumber \\
&&+\left( p+k\right) ^\mu \left\langle V\left( k,\varepsilon \right) \left| 
\bar{u}\gamma ^5b\right| B\left( p\right) \right\rangle  \label{WI1} \\
-iq_\mu \left\langle V\left( k,\varepsilon \right) \left| \bar{u}\gamma ^\mu
\gamma ^5b\right| B\left( p\right) \right\rangle &=&i\left( m_b+m_q\right)
\left\langle V\left( k,\varepsilon \right) \left| \bar{u}\gamma ^5b\right|
B\left( p\right) \right\rangle  \label{WI2}
\end{eqnarray}
Using the Ward Identity (\ref{WI1a}) in Eqs. (\ref{hcV}) and (\ref{hc2}),
and comparing the coefficient, we obtain 
\begin{equation}
F_1\left( q^2\right) =\frac{m_b+m_q}{M_B+M_V}2V\left( q^2\right)
\label{WIandF1}
\end{equation}
Again, using the Ward Identity (\ref{WI1}) in Eqs. (\ref{hc1}) and (\ref{hc3}%
), and comparing the coefficients of $\epsilon ^{\mu *}$ and that of $q^\mu $
[we do not compare the coefficients of $\left( p+k\right) ^\mu $ as heavy
quark approximation has been used in obtaining the second term of Eq. (6b)]
we obtain 
\begin{eqnarray}
F_2\left( q^2\right) &=&\frac{m_b-m_q}{M_B-M_V}A_1\left( q^2\right)
\label{WIandF2} \\
F_3\left( q^2\right) &=&-\left( m_b-m_q\right) \frac{2M_V}{q^2}\left[
A_3\left( q^2\right) -A_0\left( q^2\right) \right]  \label{WIandF3}
\end{eqnarray}

The results given in Eqs. (\ref{WIandF2}), (\ref{WIandF3}) and (\ref{WIandF1}%
) are model independent because these are derived by using Ward Identities.

In order to obtain a universal normalization of the above form factors at $%
q^2=0$, we define 
\begin{eqnarray}
\left\langle V\left( k,\epsilon \right) \left| i\bar{u}\sigma _{\alpha \beta
}b\right| B\left( p\right) \right\rangle &=&-i\varepsilon _{\alpha \beta
\rho \sigma }\varepsilon ^{*\rho }\left[ \left( p+k\right) ^\sigma
g_{+}+q^\sigma g_{-}\right] -q\cdot \varepsilon ^{*}\left( k\right)
\varepsilon _{\alpha \beta \rho \sigma }\left( p+k\right) ^\rho q^\sigma h 
\nonumber \\
&&-i\left[ q_\alpha \varepsilon _{\beta \rho \sigma \tau }\varepsilon
^{*\rho }\left( p+k\right) ^\sigma q^\tau -\alpha \leftrightarrow \beta
\right] h_1  \nonumber \\
&&-i\left[ \left( p+k\right) _\alpha \varepsilon _{\beta \rho \sigma \tau
}\varepsilon ^{*\rho }\left( p+k\right) ^\sigma q^\tau -\alpha
\leftrightarrow \beta \right] h_2  \label{VBtensor}
\end{eqnarray}

Since 
\begin{equation}
\sigma ^{\mu \nu }\gamma _5=-\frac i2\epsilon ^{\mu \nu \alpha \beta }\sigma
_{\alpha \beta },  \label{pseudoid}
\end{equation}
we can write 
\begin{equation}
\left\langle V\left( k,\epsilon \right) \left| i\bar{u}\sigma ^{\mu \nu
}\gamma _5b\right| B\left( p\right) \right\rangle =-\frac i2\varepsilon
^{\mu \nu \alpha \beta }\left\langle V\left( p^{\prime }\right) \left| i\bar{%
u}\sigma _{\alpha \beta }b\right| B\left( p\right) \right\rangle
\label{VBptensor}
\end{equation}

Using 
\[
\sigma ^{\mu \nu }=\frac i2\left[ \gamma ^\mu ,\gamma ^\nu \right] ,{ }%
\gamma ^5=i\gamma ^0\gamma ^1\gamma ^2\gamma ^3,\,\,\epsilon ^{0123}=-1 
\]
and 
\[
\epsilon ^{\mu \nu \alpha \beta }\epsilon _{\mu \delta \rho \sigma }=-\left| 
\begin{tabular}{ccccc}
$g_\delta ^\alpha $ &  & $g_\rho ^\alpha $ &  & $g_\sigma ^\alpha $ \\ 
&  &  &  &  \\ 
$g_\delta ^\beta $ &  & $g_\rho ^\beta $ &  & $g_\sigma ^\beta $ \\ 
&  &  &  &  \\ 
$g_\delta ^\nu $ &  & $g_\rho ^\nu $ &  & $g_\sigma ^\nu $%
\end{tabular}
\right| 
\]
we obtain from Eq. (\ref{VBtensor}) 
\begin{eqnarray}
\left\langle V\left( k,\epsilon \right) \left| i\bar{u}\sigma ^{\mu \nu
}q_\nu \gamma _5b\right| B\left( p\right) \right\rangle &=&\varepsilon
^{*\mu }\left[ \left( M_B^2-M_V^2\right) g_{+}+q^2g_{-}\right]  \nonumber \\
&&-q\cdot \varepsilon ^{*}\left( k\right) \left[ \left( p+k\right) ^\mu
g_{+}+q^\mu g_{-}\right]  \nonumber \\
&&+q\cdot \varepsilon ^{*}\left( k\right) \left[ q^2\left( p+k\right) ^\mu
-\left( M_B^2-M_V^2\right) q^\mu \right] h  \nonumber \\
&&-\varepsilon ^{*\mu }\left( M_B^2-M_V^2\right) ^2h_2  \nonumber \\
&&+\varepsilon ^{*\mu }\left( 2M_B^2+2M_V^2-q^2\right) q^2h_2  \nonumber \\
&&+q\cdot \varepsilon ^{*}\left( k\right) \left( p+k\right) ^\mu \left(
M_B^2-M_V^2-q^2\right) h_2  \nonumber \\
&&+q\cdot \varepsilon ^{*}\left( k\right) q^\mu \left(
-M_B^2-3M_V^2+q^2\right) h_2  \label{VBptensorqnu}
\end{eqnarray}
Equating Eqs. (\ref{hc3}) and (\ref{VBptensorqnu}) and comparing various
coefficients 
\begin{eqnarray}
F_3\left( q^2\right) &=&-g_{-}-\left( M_B^2-M_V^2\right) h-\left(
M_B^2+3M_V^2-q^2\right) h_2  \label{F3qsq} \\
F_2\left( q^2\right) &=&g_{+}-q^2h-\left( M_B^2-M_V^2-q^2\right) h_2-\frac{%
q^2}{M_B^2-M_V^2}F_3\left( q^2\right)  \nonumber \\
&=&g_{+}+\frac{q^2}{M_B^2-M_V^2}g_{-}-\left[ M_B^2-M_V^2-\frac{q^2}{%
M_B^2-M_V^2}\left( 2M_B^2+2M_V^2-q^2\right) \right] h_2  \nonumber \\
&&  \label{F2qsq}
\end{eqnarray}
Further, from Eq. (\ref{VBtensor}) 
\begin{equation}
\left\langle V\left( k,\epsilon \right) \left| i\bar{u}\sigma _{\alpha \beta
}q^\beta b\right| B\left( p\right) \right\rangle =-i\varepsilon _{\alpha
\beta \rho \sigma }\varepsilon ^{*\rho }p^\sigma q^\beta 2\left[
g_{+}-q^2h_1-\left( M_B^2-M_V^2\right) h_2\right]  \label{VBtensor1}
\end{equation}
Comparing Eqs. (\ref{VBtensor1}) and (\ref{hc2}), we obtain 
\begin{equation}
F_1\left( q^2\right) =2\left[ g_{+}-q^2h_1-\left( M_B^2-M_V^2\right)
h_2\right]  \label{F1qsq}
\end{equation}
Note that above relations ensure that $F_2\left( 0\right) =\left( 1/2\right)
F_1\left( 0\right) $. Equation (\ref{WIandF2}), by using the value of $%
F_2\left( q^2\right) $, becomes 
\begin{eqnarray}
A_1\left( q^2\right) &=&\frac{M_B-M_V}{m_b-m_q}\left\{ g_{+}+\frac{q^2}{%
M_B^2-M_V^2}g_{-}\right.  \nonumber \\
&&\left. -\left[ M_B^2-M_V^2-\frac{q^2}{M_B^2-M_V^2}\left(
2M_B^2+2M_V^2-q^2\right) \right] h_2\right\}  \label{A1qsq1}
\end{eqnarray}
Equation (\ref{F1qsq}), by using the value of $F_1\left( q^2\right) $ given
in Eq. (\ref{WIandF1}), becomes 
\begin{equation}
V\left( q^2\right) =\frac{M_B+M_V}{m_b+m_q}\left[ g_{+}-q^2h_1-\left(
M_B^2-M_V^2\right) h_2\right]  \label{Vqsq1}
\end{equation}
Rearranging Eq. (\ref{A3qsq}), we get 
\begin{eqnarray}
A_2\left( q^2\right) &=&\frac{M_B+M_V}{M_B-M_V}A_1\left( q^2\right) -\frac{%
2M_V}{M_B-M_V}A_3\left( q^2\right)  \nonumber \\
&=&\frac{M_B+M_V}{M_B-M_V}A_1\left( q^2\right) -\frac{2M_V}{M_B-M_V}\left[
A_3\left( q^2\right) -A_0\left( q^2\right) \right] -\frac{2M_V}{M_B-M_V}%
A_0\left( q^2\right)  \nonumber \\
&=&\frac{M_B+M_V}{m_b-m_q}\left\{ g_{+}-q^2h-\left[ M_B^2-M_V^2+\frac{q^2}{%
M_B^2-M_V^2}\left( 3M_B^2+5M_V^2-2q^2\right) \right] h_2\right\}  \nonumber
\\
&&-\frac{2M_V}{M_B-M_V}A_0\left( q^2\right)  \label{A2qsq1}
\end{eqnarray}
where we have used Eqs. (\ref{WIandF3}) and (\ref{F3qsq}).

The merit of above relations is that the normalization of the form factors $%
A_1$, $V$, $F_1$ and $F_2$ at $q^2=0$ is determined by a single constant $%
\left[ g_{+}\left( 0\right) -\left( M_B^2-M_V^2\right) h_2\left( 0\right)
\right] $ while those of $A_2$ and $A_0$ are determined by $\left[
g_{+}\left( 0\right) -\left( M_B^2-M_V^2\right) h_2\left( 0\right) \right] $
and $A_0\left( 0\right) $.

\section{Pole contribution\label{pcanddff}}

Only $h_1$, $g_{-}$, $h$, and $A_0$ get pole contributions from $B^{*}\left(
1^{-}\right) $, $B_A^{*}\left( 1^{+}\right) $, and $B\left( 0^{-}\right) $
mesons. These are given by 
\begin{eqnarray}
\left. h_1\right| _{{pole}} &=&-\frac 12\frac{g_{B^{*}B\rho ^{-}}}{%
M_{B^{*}}^2}\frac{f_T^{B^{*}}}{1-q^2/M_{B^{*}}^2}=\frac{R_V}{M_{B^{*}}^2}%
\frac 1{1-q^2/M_{B^{*}}^2} \\
\left. g_{-}\right| _{{pole}} &=&-\frac{g_{B_A^{*}B\rho ^{-}}}{%
M_{B_A^{*}}^2}\frac{f_T^{B_A^{*}}}{1-q^2/M_{B_A^{*}}^2}=\frac{R_A^S}{%
M_{B_A^{*}}^2}\frac 1{1-q^2/M_{B_A^{*}}^2} \\
\left. h\right| _{{pole}} &=&\frac 12\frac{f_{B_AB\rho ^{-}}}{%
M_{B_A^{*}}^2}\frac{f_T^{B_A^{*}}}{1-q^2/M_{B_A^{*}}^2}=\frac{R_A^D}{%
M_{B_A^{*}}^2}\frac 1{1-q^2/M_{B_A^{*}}^2} \\
\left. A_0\left( q^2\right) \right| _{{pole}} &=&\frac{g_{BB\rho ^{-}}}{%
M_\rho }f_B\frac{q^2/M_B^2}{1-q^2/M_B^2}=R_0\frac{q^2/M_B^2}{1-q^2/M_B^2}
\end{eqnarray}

On the other hand, only $g_{+}$, $g_{-}$, and $A_0$ get contribution from
quark $\triangle $--graph. Therefore we shall put $h_2=0$ in what follows.

The above coupling constants are defined as follows: 
\begin{eqnarray}
\left\langle B^{*+}\left( \eta ,q\right) \rho ^{-}\left( \epsilon ,k\right)
\left| B^0\left( p\right) \right. \right\rangle &=&ig_{B^{*}B\rho
^{-}}\epsilon _{\alpha \rho \mu \sigma }\epsilon ^{*\alpha }q^\rho p^\sigma
\eta ^{*\mu }  \nonumber \\
\left\langle 0\left| i\bar{u}\sigma _{\mu \nu }b\right| B^{*}\left( \eta
,q\right) \right\rangle &=&f_T^{B^{*}}\left( q_\mu \eta _\nu -q_\nu \eta
_\mu \right)  \label{cc2} \\
\left\langle B_A^{*+}\left( \eta ,q\right) \rho ^{-}\left( \epsilon
,k\right) \left| B^0\left( p\right) \right. \right\rangle
&=&ig_{B_A^{*}B\rho ^{-}}\epsilon ^{*}\cdot \eta ^{*}-f_{B_A^{*}B\rho
^{-}}q\cdot \epsilon ^{*}k\cdot \eta ^{*}  \nonumber \\
\left\langle 0\left| i\bar{u}\sigma _{\mu \nu }b\right| B^{*}\left( \eta
,q\right) \right\rangle &=&f_T^{B_A^{*}}\epsilon _{\mu \nu \alpha \beta
}\eta ^\alpha q^\beta  \nonumber
\end{eqnarray}
Now if in the Ward Identity (\ref{WI1a}), we take the matrix elements
between $\left\langle 0\right| $ and $\left| B^{*}\right\rangle $, we obtain 
\begin{equation}
\left\langle 0\left| i\bar{u}\sigma ^{\mu \nu }q_\nu b\right| B^{*}\left(
\eta ,q\right) \right\rangle =-\left( m_b+m_q\right) f_{B^{*}}\eta ^\mu
\end{equation}
where $\left\langle 0\left| i\bar{u}\gamma ^\mu b\right| B^{*}\left( \eta
,q\right) \right\rangle =f_{B^{*}}\eta ^\mu $. Thus using Eq. (\ref{cc2}),
we obtain 
\begin{equation}
f_T^{B^{*}}=\frac{m_b+m_q}{M_{B^{*}}^2}f_{B^{*}}=\frac{m_b+m_q}{M_{B^{*}}}%
f_B=\frac{M_B}{M_{B^{*}}}f_B=f_B
\end{equation}
It is easy to see that same relation is obtained if one uses heavy quark
spin symmetry. Similarly, if in the Ward Identity (\ref{WI1}), we take the
matrix elements between $\left\langle 0\right| $ and $\left|
B_A^{*}\right\rangle $, we obtain on using that $\left\langle 0\left| i\bar{u%
}\gamma _5b\right| B_A^{*}\left( \eta ,q\right) \right\rangle =0,$%
\begin{equation}
\left\langle 0\left| i\bar{u}\sigma ^{\mu \nu }q_\nu \gamma _5b\right|
B_A^{*}\left( \eta ,q\right) \right\rangle =\left( m_b-m_q\right)
f_{B_A^{*}}\eta ^\mu ,
\end{equation}
where $\left\langle 0\left| i\bar{u}\gamma ^\mu \gamma _5b\right|
B_A^{*}\left( \eta ,q\right) \right\rangle =f_{B_A^{*}}\eta ^\mu $. By using
the identity (\ref{pseudoid}), we obtain 
\begin{equation}
f_T^{B_A^{*}}=-\frac{m_b-m_q}{M_{B_A^{*}}^2}f_{B_A^{*}}
\end{equation}

\section{Form factors and determination of coupling constants \label{ffandcc}
}

The form factors $V\left( q^2\right) $ [Eq. (\ref{Vqsq1})], $A_1\left(
q^2\right) $ [Eq. (\ref{A1qsq1})], and $A_2\left( q^2\right) $ [Eq. (\ref
{A2qsq1})] can be written as 
\begin{eqnarray}
V\left( q^2\right) &=&\frac{M_B+M_V}{m_b+m_q}\left\{ g_{+}\left( q^2\right)
-q^2h_1\left( q^2\right) \right\}  \nonumber \\
&\simeq &\frac{M_B+M_V}{M_B}\left\{ g_{+}\left( q^2\right) -R_V\frac{q^2}{%
M_{B^{*}}^2}\frac 1{1-q^2/M_{B^{*}}^2}\right\}  \label{Vqsq2} \\
A_1\left( q^2\right) &=&\frac{M_B-M_V}{m_b-m_q}\left\{ g_{+}\left(
q^2\right) +\frac{q^2}{M_B^2-M_V^2}g_{-}\left( q^2\right) \right\}  \nonumber
\\
&=&\frac{M_B-M_V}{m_b-m_q}\left\{ g_{+}\left( q^2\right) +\frac{q^2}{%
M_B^2-M_V^2}\left( \tilde{g}_{-}\left( q^2\right) +\left. g_{-}\left(
q^2\right) \right| _{{pole}}\right) \right\}  \nonumber \\
&\simeq &g_{+}\left( q^2\right) +\frac{q^2}{M_B^2-M_V^2}\tilde{g}_{-}\left(
q^2\right) +\frac{R_A^S}{M_B^2-M_V^2}\frac{q^2}{M_{B_A^{*}}^2}\frac
1{1-q^2/M_{B_A^{*}}^2}  \label{A1qsq2} \\
A_2\left( q^2\right) &=&\frac{M_B+M_V}{m_b-m_q}\left\{ g_{+}\left(
q^2\right) -q^2h\left( q^2\right) \right\} -\frac{2M_V}{M_B-M_V}A_0\left(
q^2\right)  \nonumber \\
&\simeq &\frac{M_B+M_V}{M_B-M_V}\left\{ g_{+}\left( q^2\right) -R_A^D\frac{%
q^2}{M_{B_A^{*}}^2}\frac 1{1-q^2/M_{B_A^{*}}^2}\right\} -\frac{2M_V}{M_B-M_V}%
A_0\left( q^2\right)  \label{A2qsq2}
\end{eqnarray}
The behaviour of $g_{+}\left( q^2\right) $, $\tilde{g}_{-}\left( q^2\right) $
and $A_0\left( q^2\right) $ near $q^2\rightarrow 0$ is known from LEET 
\begin{eqnarray}
g_{+}\left( q^2\right) &=&\frac{\xi _{\perp }\left( 0\right) }{\left(
1-q^2/M_B^2\right) ^2}=-\tilde{g}_{-}\left( q^2\right)  \label{gpqsq} \\
A_0\left( q^2\right) &=&\left( 1-\frac{M_V^2}{M_BE_V}\right) \xi _{\parallel
}\left( q^2\right) +\frac{M_V}{M_B}\xi _{\perp }\left( q^2\right)
\label{A0qsq} \\
A_0\left( 0\right) &=&\frac{M_B^2-M_V^2}{M_B^2+M_V^2}\xi _{\parallel }\left(
0\right) +\frac{M_V}{M_B}\xi _{\perp }\left( 0\right)  \label{A00} \\
E_V &=&\frac{M_B}2\left( 1-\frac{q^2}{M_B^2}+\frac{M_V^2}{M_B^2}\right) .
\label{EV}
\end{eqnarray}

The pole terms in the relations (\ref{Vqsq2}), (\ref{A1qsq2}) and (\ref
{A2qsq2}) are expected to dominate near $q^2=M_{B^{*}}^2$ or $M_{B_A^{*}}^2$%
. Therefore, we cannot expect these relations, obtained from Ward
identities, to hold for all $q^2$ for which we use the parametarization,
suggested by above behavior, near $q^2=0$ and near the pole 
\begin{equation}
F\left( q^2\right) =\frac{F\left( 0\right) }{\left( 1-q^2/M^2\right) \left(
1-q^2/M^{\prime 2}\right) },  \label{cFqsq}
\end{equation}
where $M^2$ is $M_{B^{*}}^2$ or $M_{B_A^{*}}^2$ and $M^{\prime }$ is the
radial excitation of $M$. This parameterization takes into account potential
corrections to single pole dominance, presumably arising from radial
excitations of M \cite{ZPC48-55}, as suggested by dispersion relations \cite
{PLB214-459,ZPC41-217} (a detailed discussion on this point one can find in 
\cite{ZPC48-55}). It also takes care of off-mass-shell-ness of couplings of $%
B^{*}$ or $B_A^{*}$ with $B\rho $ channel \cite{ZPC48-55} by replacing the
pole contribution 
\begin{equation}
\frac R{1-q^2/M^2}{\,\, with\,\, }\frac{R\left( 1-M^2/M^{\prime 2}\right) }{%
\left( 1-q^2/M^2\right) \left( 1-q^2/M^{\prime 2}\right) }{.}
\label{replace}
\end{equation}
Since $g_{+}\left( q^2\right) $ and $\tilde{g}_{-}\left( q^2\right) $ do not
have a pole at $q^2=M_{B^{*}}^2$, we get 
\[
V\left( q^2\right) \left. \left( 1-q^2/M_{B^{*}}^2\right) \right|
_{q^2=M_{B^{*}}^2}=-\frac{M_B+M_V}{M_B}R_V{.} 
\]

This becomes 
\begin{equation}
R_V\equiv -\frac 12g_{B^{*}B\rho ^{-}}f_T^{B^{*}}=-\frac 12g_{B^{*}B\rho
^{-}}f_B=-\frac{g_{+}\left( 0\right) }{1-M_{B^{*}}^2/M_{B^{*}}^{\prime 2}}
\label{RV}
\end{equation}

Similarly, using parameterization (\ref{cFqsq}) for the first term on the
right hand side of Eq. (\ref{A2qsq2}), 
\begin{equation}
R_A^D\equiv -\frac 12f_{B_A^{*}B\rho ^{-}}f_T^{B_A^{*}}=\frac
12f_{B_A^{*}B\rho ^{-}}\frac{M_B-M_V}{M_{B_A^{*}}^2}f_{B_A^{*}}=-\frac{%
g_{+}\left( 0\right) }{1-M_{B_A^{*}}^2/M_{B_A^{*}}^{\prime 2}}  \label{RAD}
\end{equation}
We cannot use the parameterization (\ref{cFqsq}) for $A_1\left( q^2\right) $
since near $q^2=0$, $A_1\left( q^2\right) $ behaves as $g_{+}\left(
q^2\right) \left[ 1-q^2/\left( M_B^2-M_V^2\right) \right] $. This suggests
the following 
\begin{equation}
A_1\left( q^2\right) =\frac{g_{+}\left( 0\right) }{\left(
1-q^2/M_{B_A^{*}}^2\right) \left( 1-q^2/M_{B_A^{*}}^{\prime 2}\right) }%
\left( 1-\frac{q^2}{M_B^2-M_V^2}\right)  \label{A1qsq3}
\end{equation}
so that, since $g_{+}$ and $\tilde{g}_{-}$ do not have pole at $%
q^2=M_{B_A^{*}}^2$, 
\begin{eqnarray}
&&\left. \left. A_1\left( q^2\right) \left[ 1-q^2/M_{B_A^{*}}^2\right]
\right| _{q^2=M_{B_A^{*}}^2}=g_{+}\left( 0\right) \frac{1-M_{B_A^{*}}^2/%
\left( M_B^2-M_V^2\right) }{1-M_{B_A^{*}}^2/M_{B_A^{*}}^{\prime 2}}\right. 
\nonumber \\
&&\left. =\frac{R_A^s}{M_B^2-M_V^2}M_{B_A^{*}}^2=-g_{B_A^{*}B\rho ^{-}}\frac{%
M_B-M_V}{M_{B_A^{*}}^2}f_{B_A^{*}}\frac 1{M_B^2-M_V^2}\right.  \label{A1qsq4}
\end{eqnarray}
Most of the estimates of $g_{+}\left( 0\right) $ suggest that \cite{gplus0} 
\begin{equation}
g_{+}\left( 0\right) =0.29\pm 0.04\,{\cite{PRD58-094016}}  \label{aaaaa1}
\end{equation}
\begin{table}[tbp]
\caption{$B$-mesons masses in GeV \protect\cite{ZPC29-637,PRD21-203} }
\begin{center}
$
\begin{tabular}{|c|c|c|c|}
\hline
& $J^P$ & $M$ & $M^{\prime }/M$ \\ \hline
$M_B$ & $0^{-}$ & $5.28$ & $1.14$ \\ \hline
$M_{B^{*}}$ & $1^{-}$ & $5.33$ & $1.14$ \\ \hline
$M_{B_A^{*}}$ & $1^{+}$ & $5.71$ & $1.12$ \\ \hline
\end{tabular}
$
\end{center}
\end{table}

For numerical work we shall use the $B$-meson masses given in Table 1. Thus
we have the prediction from Eq. (\ref{RV}) 
\begin{equation}
g_{B^{*}B\rho ^{-}}=\frac{2\left( 1.26\pm 0.17\right) }{f_B}  \label{gBsBrho}
\end{equation}
This prediction [$f_B=180$ MeV] is consistant with its various estimates
e.g. QCD sum rules give it $\sqrt{2}\left( 11\right) $ GeV$^{-1}$ with $%
g_{B^{*}B\rho }=\sqrt{2}g_{B^{*}B\rho ^0}$ (where $g_{B^{*}B\rho ^0}$ has
been estimated in \cite{ZPC41-217}) and the constituent quark model (CQM)
predicts it to be \cite{PLB214-459} 
\begin{equation}
\sqrt{2}\left\{ \frac 12\left( \frac 1{m_b}+\frac 1{m_u}\right) \frac{M_\rho
^2}{f_{\rho ^{-}}}\sqrt{2}\right\} \simeq \sqrt{2}\left( 9.75\right) \,{
GeV}^{-1}.  \label{gBsBrhoCQM}
\end{equation}
The predictions (\ref{RAD}) and (\ref{A1qsq4}) cannot be tested at present
since we do not have any information on $f_{B_A^{*}}$. However, the ratio of 
$S$-wave and $D$-wave couplings is predicted to be 
\begin{eqnarray}
\frac{g_{B_A^{*}B\rho ^{-}}}{f_{B_A^{*}B\rho ^{-}}} &=&\frac{%
M_{B_A^{*}}^2-\left( M_B^2-M_V^2\right) }2  \nonumber \\
g_{B_A^{*}B\rho ^{-}} &=&2.66\times f_{B_A^{*}B\rho ^{-}}\,{ GeV}^2
\label{swave-Dwave}
\end{eqnarray}

The different values of $F\left( 0\right) $'s are 
\begin{eqnarray}
V\left( 0\right) &=&\frac{M_B+M_V}{M_B}g_{+}\left( 0\right) \\
A_1\left( 0\right) &=&g_{+}\left( 0\right) \\
A_2\left( 0\right) &=&\frac{M_B+M_V}{M_B-M_V}g_{+}\left( 0\right) -\frac{2M_V%
}{M_B-M_V}A_0\left( 0\right)
\end{eqnarray}
where $g_{+}\left( 0\right) $ is given by (\ref{aaaaa1}). As far as $%
A_0\left( 0\right) $ is concerned we can get some infor mation by writing 
\begin{eqnarray}
A_0\left( q^2\right) &=&\tilde{A}_0\left( q^2\right) +\left. A_0\left(
q^2\right) \right| _{{pole}}  \nonumber \\
&=&\tilde{A}_0\left( q^2\right) +R_0\frac{q^2/M_B^2\left(
1-M_B^2/M_B^{\prime 2}\right) }{\left( 1-q^2/M_B^2\right) \left(
1-q^2/M_B^{\prime 2}\right) }
\end{eqnarray}
Using the parameterization for $A_0\left( q^2\right) $ similar to (\ref
{cFqsq}) and the same argument as before [cf. Eq. (\ref{RV})], we get 
\begin{eqnarray}
A_0\left( 0\right) &=&\tilde{A}_0\left( 0\right) =R_0\left( 1-\frac{M_B^2}{%
M_B^{\prime 2}}\right)  \nonumber \\
&=&\frac{g_{BB\rho ^{-}}}{M_\rho }f_B\left( 1-\frac{M_B^2}{M_B^{\prime 2}}%
\right)
\end{eqnarray}
Using the $\rho $-dominance 
\begin{equation}
g_{BB\rho ^{-}}=\frac{M_\rho ^2}{f_{\rho ^{-}}}\frac 1{F_\rho \left(
0\right) }
\end{equation}
so that 
\begin{equation}
A_0\left( 0\right) =f_B\frac{M_\rho }{f_{\rho ^{-}}}\left( 1-\frac{M_B^2}{%
M_B^{\prime 2}}\right) \frac 1{F_\rho \left( 0\right) }
\end{equation}
where $F_\rho \left( 0\right) $ takes care off mass shell-ness of $\rho $%
-meson and is given by \cite{PLB214-459} 
\begin{equation}
F_\rho \left( 0\right) =\left( \frac{g_{\rho \pi \pi }f_{\rho ^{-}}}{\sqrt{2}%
M_\rho ^2}\right) _{{exp}}^{-1}=0.80,
\end{equation}
where we have used 
\begin{equation}
\left( \frac{g_{\rho \pi \pi }f_{\rho ^{-}}}{\sqrt{2}M_\rho ^2}\right) _{%
{exp}}=1.22\pm 0.03.
\end{equation}
Using $f_B=180$ MeV, $f_{\rho ^{-}}/M_\rho =198$ MeV, 
\begin{equation}
A_0\left( 0\right) =0.26.  \label{A00a}
\end{equation}
Although LEET does not give any relationship between $\xi _{\parallel
}\left( 0\right) $ and $\xi _{\perp }\left( 0\right) $ but due to some
numerical coincidence in the LCSR expressions for $\xi _{\parallel }$ and $%
\xi _{\perp }$ \cite{PRD60-014001,dgplus0} 
\begin{equation}
\xi _{\parallel }\left( 0\right) \simeq \xi _{\perp }\left( 0\right)
=g_{+}\left( 0\right)  \label{xillandperp}
\end{equation}
so that, from Eq. (\ref{A00}) 
\begin{equation}
A_0\left( 0\right) =1.10{ }g_{+}\left( 0\right) \simeq 0.320\pm 0.044
\label{A00b}
\end{equation}
consistent with the above estimate (\ref{A00a}). Another estimate of $%
A_0\left( 0\right) $ in light-cone sum rules (LCSR) approach is 0.372.

We summarise the form factors which we shall use for our numerical work [$%
M_V=M_\rho =0.770$ GeV] 
\label{ffvalues}
\begin{eqnarray}
V\left( q^2\right) &=&\frac{V\left( 0\right) }{\left( 1-q^2/M_B^2\right)
\left( 1-q^2/M_B^{\prime 2}\right) }  \label{Vqsqp} \\
A_1\left( q^2\right) &=&\frac{A_1\left( 0\right) }{\left(
1-q^2/M_{B_A^{*}}^2\right) \left( 1-q^2/M_{B_A^{*}}^{\prime 2}\right) }%
\left( 1-\frac{q^2}{M_B^2-M_V^2}\right)  \label{A1qsqp} \\
A_2\left( q^2\right) &=&\frac{\tilde{A}_2\left( 0\right) }{\left(
1-q^2/M_{B_A^{*}}^2\right) \left( 1-q^2/M_{B_A^{*}}^{\prime 2}\right) }-%
\frac{2M_V}{M_B-M_V}\frac{A\left( 0\right) }{\left( 1-q^2/M_B^2\right)
\left( 1-q^2/M_B^{\prime 2}\right) }  \label{A2qsqp}
\end{eqnarray}
where 
\begin{equation}
\begin{tabular}{ccc}
$V\left( 0\right) $ & $=$ & $0.332\pm 0.046,$ \\ 
$A_1\left( 0\right) $ & $=$ & $0.29\pm 0.04,$ \\ 
$\tilde{A}_2\left( 0\right) $ & $=$ & $0.389\pm 0.054$.
\end{tabular}
\label{ff0values}
\end{equation}
These form factors (\ref{ffvalues}) are plotted in Fig. \ref{figva1c}, where
the comparison of the form factor vs. lattice data \cite
{PLB416-392,BaBar,lat9611016} and calculations within different approaches
is given. Solid line--our result, dotted line--LCSR \cite{PRD58-094016},
dashed line--quark model \cite{PLB436-344}, dash-dotted line--lattice
constrained parametriation of \cite{PLB416-392}.

\section{Decay distribution\label{sec-decay-distribution}}

Four independent kinematical variables completely describe the semileptonic
decay $B\rightarrow \rho l\bar{\nu}$, where the vector meson decays to two
pseudoscalar mesons, $\rho \rightarrow \pi \pi $. The four variables
commonly used are $q^2=\left( p-k\right) ^2$ and the three angles ($\theta
_l $, $\theta _\rho $, and $\chi $). The angle $\theta _l$ is measured in
the $W^{*}$ (or $l\bar{\nu}$) rest frame, it is the polar angle between the
charged lepton and the direction opposite to that of vector meson. The angle 
$\theta _\rho $ is measured in the rest frame of vector meson, where the
pseudoscalar mesons are back to back, and it is the polar angle between one
of these mesons. The angle $\chi $ is chosen to be azimuthal angle between
the plane of flight of lepton and of pseudoscalar mesons.

The differential decay rate for $B\rightarrow \rho l\nu $, $\rho \rightarrow
\pi \pi $ can be expressed in terms of these four variables $q^2$, $\theta
_l $, $\theta _\rho $, and $\chi $ \cite{PRD41-142,ZPC46-93} 
\begin{eqnarray}
\frac{d\Gamma \left( B\rightarrow \rho l\nu {, }\rho \rightarrow \pi
\pi \right) }{dq^2\,d\cos \theta _\rho \,d\cos \theta _l\,d\chi } &=&\frac
3{8\left( 4\pi \right) ^4}G_F^2\left| V_{ub}\right| ^2\frac{\left| {\bf k}%
_\rho \right| q^2}{M_B^2}{\cal B}\left( \rho \rightarrow \pi \pi \right) 
\nonumber \\
&&\times \left\{ \left( 1-\cos \theta _l\right) ^2\sin ^2\theta _\rho \left|
H_{+}\left( q^2\right) \right| ^2\right.  \nonumber \\
&&+\left( 1+\cos \theta _l\right) ^2\sin ^2\theta _\rho \left| H_{-}\left(
q^2\right) \right| ^2+4\sin ^2\theta _l\cos ^2\theta _\rho \left| H_0\left(
q^2\right) \right| ^2  \nonumber \\
&&-4\sin \theta _l\left( 1-\cos \theta _l\right) \sin \theta _\rho \cos
\theta _\rho \cos \chi H_{+}\left( q^2\right) H_0\left( q^2\right)  \nonumber
\\
&&+4\sin \theta _l\left( 1+\cos \theta _l\right) \sin \theta _\rho \cos
\theta _\rho \cos \chi H_{-}\left( q^2\right) H_0\left( q^2\right)  \nonumber
\\
&&\left. -2\sin ^2\theta _l\sin ^2\theta _\rho \cos 2\chi H_{+}\left(
q^2\right) H_{-}\left( q^2\right) \right\}  \label{diff-decay}
\end{eqnarray}
where 
\begin{equation}
\left| {\bf k}_\rho \right| =\sqrt{\frac{\left( M_B^2+M_\rho ^2-q^2\right) ^2%
}{4M_B^2}-M_\rho ^2}
\end{equation}
is the magnitude of three-momentum of $\rho $ in the rest frame of $B$ and
is a function of $q^2$.

Bescause the parent meson has spin zero, the vector meson and the $W^{*}$
must have the same helicity. The amplitudes for the helicities 0, +1, and
--1 are proportional to $H_0\left( q^2\right) $, $H_{+}\left( q^2\right) $,
and $H_{-}\left( q^2\right) $. The detailed dynamics of the hadronic current
are described by the variation of these helicity amplitudes with $q^2$.
Equation (\ref{diff-decay}) incorporates the $V-A$ structure of the leptonic
current, as well as the assumption that the mass of the charged lepton can
be neglected.

The helicity amplitudes can in turn be related to the two axial form
factors, $A_1\left( q^2\right) $ and $A_2\left( q^2\right) $, and the vector
form factor $V\left( q^2\right) $, which appear in the hadronic current (\ref
{hc1}) and (\ref{hcV}) respectively, 
\begin{equation}
H_0\left( q^2\right) =\frac 1{2M_\rho \sqrt{q^2}}\left[ \left(
M_B^2-q^2-M_\rho ^2\right) \left( M_B+M_\rho \right) A_1\left( q^2\right) -%
\frac{4M_B^2\left| {\bf k}_\rho \right| ^2}{M_B+M_\rho }A_2\left( q^2\right)
\right]  \label{cH0}
\end{equation}
and 
\begin{equation}
H_{\pm }\left( q^2\right) =\left( M_B+M_\rho \right) A_1\left( q^2\right)
\mp \frac{2M_B\left| {\bf k}_\rho \right| }{M_B+M_\rho }V\left( q^2\right)
\label{cHpm}
\end{equation}
The form factors $A_1\left( q^2\right) $, $A_2\left( q^2\right) $, and $%
V\left( q^2\right) $ are dimensionless and relatively real, since CP is
conserved in these decays and there are no final state strong interactions 
\cite{ZPC38-511}. We note that, while $A_2$ contributes only to $H_0$ and $V$
contributes only to $H_{\pm }$, $A_1$ contributes to all three helicity
amplitudes. At high $q^2$ (small $\left| {\bf k}_\rho \right| $), each of
the helicity amplitudes is dominated by $A_1$. At $q^2=0$, the over all
factor of $q^2$ in (\ref{diff-decay}) causes all the contributions to the
decay rate to vanish except that from $\left| H_0\left( q^2\right) \right|
^2 $, since $H_0$ contains a factor of $1/\sqrt{q^2}$ and at $q^2=q_{\max
}^2=\left( M_B-M_\rho \right) ^2$, the over all factor of $\left| {\bf k}%
_\rho \right| $ vanishes, causing the rate to do likewise.

Integrating (\ref{diff-decay}) over the angles $\theta _\rho $ and $\chi $,
we obtain 
\begin{eqnarray}
\frac{d^2\Gamma }{dq^2dE_l} &=&\frac{G_F^2\left| V_{ub}\right| ^2}{128\pi ^3}%
\frac{q^2}{M_B^2}\left\{ \left( 1+\cos \theta _l\right) ^2\left| H_{-}\left(
q^2\right) \right| ^2\right.  \nonumber \\
&&\left. +\left( 1-\cos \theta _l\right) ^2\left| H_{+}\left( q^2\right)
\right| ^2+2\left( 1-\cos ^2\theta _l\right) \left| H_0\left( q^2\right)
\right| ^2\right\}  \label{doublediff}
\end{eqnarray}
where $\theta _l$ is the polar angle between the $\rho $ and the lepton $l$
in the $\left( l\bar{\nu}_l\right) $ center of mass system, and 
\begin{equation}
2M_B\left| {\bf k}_\rho \right| \cos \theta _l=M_B^2-M_\rho ^2+q^2-4M_BE_l%
{.}  \label{El-costhl}
\end{equation}

The $E_l$ (or $\cos \theta _l$) integration of Eq. (\ref{doublediff}) can be
done trivially and results in the differential $q^2$-distribution 
\begin{equation}
\frac{d\Gamma }{dq^2}=\frac{G_F^2\left| V_{ub}\right| ^2}{96\pi ^3}\left| 
{\bf k}_\rho \right| \frac{q^2}{M_B^2}{\cal B}\left( \rho \rightarrow \pi
\pi \right) \left\{ \left| H_{-}\left( q^2\right) \right| ^2+\left|
H_{+}\left( q^2\right) \right| ^2+\left| H_0\left( q^2\right) \right|
^2\right\}  \label{ddhelicitysq}
\end{equation}

By numerically integrating Eq. (\ref{doublediff}) over lepton energy $E_l$
in the interval $0\leq E_l\leq 2.6$ and over $q^2$ in the interval 
\[
0\leq q^2\leq 2E_l\left( \frac{M_B^2-M_\rho ^2-2M_BE_l}{M_B-2E_l}\right) 
{,} 
\]
the width of $B\rightarrow \rho l\nu $ turns out to be 
\begin{equation}
\Gamma \left( B\rightarrow \rho l\nu \right) =15.82\pm 2.18{ }\left|
V_{ub}\right| ^2{ ps}^{-1}{,}  \label{width-our}
\end{equation}
where we have used the form factors given in Eqs. (\ref{ffvalues}).

The branching ratio of $B^0\rightarrow \rho ^{-}l^{+}\nu $ is measured by
CLEO\ collaboration \cite{PRD61-052001}, ${\cal B}_{{expt}%
}(B^0\rightarrow \rho ^{-}l^{+}\nu )=\left( 2.57\pm 0.29_{-0.46}^{+0.33}\pm
0.41\right) \times 10^{-4}$, where the errors are statistical, systematic,
and theoretical. The CKM matrix element $\left| V_{ub}\right| $ can be
obtained from the branching ratio ${\cal B}_{{expt}}\left(
B^0\rightarrow \rho ^{-}l^{+}\nu \right) $ using 
\begin{eqnarray}
\left| V_{ub}\right| &=&\sqrt{\frac{{\cal B}_{{expt}}\left(
B^0\rightarrow \rho ^{-}l^{+}\nu \right) }{\Gamma _{{thy}}\left(
B^0\rightarrow \rho ^{-}l^{+}\nu \right) \,\,\tau _{B^0}}}  \nonumber \\
&=&3.24\times 10^{-3}{.}  \label{Vubour}
\end{eqnarray}
We use $\tau _{B^0}=1.548\pm 0.032$ ps \cite{PRD66-010001}. Recently, the
BaBar Collaboration \cite{ex0301001} predicted the branching ratio ${\cal B}%
_{{expt}}(B^0\rightarrow \rho ^{-}l^{+}\nu )=\left( 3.29\pm 0.42\pm
0.47\pm 0.60\right) \times 10^{-4}$ and $\left| V_{ub}\right| =\left(
3.64\pm 0.22\pm 0.25_{-0.56}^{+0.39}\right) \times 10^{-3}$, where the
uncertainities are statistical, systematic, and theoretical respectively. By
using the branching ratio predicted by BaBar, and decay width (\ref
{width-our}), we obtain 
\begin{equation}
\left| V_{ub}\right| =3.67\times 10^{-3}{.}  \label{Vubour2}
\end{equation}
Another recent study by CLEO \cite{ex0304019}, gave the branching ratio $%
{\cal B}\left( B^0\rightarrow \rho ^{-}l^{+}\nu \right) =(2.17\pm
0.34_{-0.54}^{+0.47}\pm 0.41\pm 0.01)\times 10^{-4}$, $\left| V_{ub}\right|
=\left( 3.00\pm 0.21_{-0.35}^{+0.29}\,_{-0.38}^{+0.49}\pm 0.28\right) \times
10^{-3}$, where the errors in CLEO\ analyses are statistical, experimental
systematic, theoretical systematicbased on lattice QCD and LCSR
uncertainities, and $\rho l\nu $ form factor shape uncertainity,
respectively. By using this branching ratio given by CLEO \cite{ex0304019},
we obtain 
\[
\left| V_{ub}\right| =2.98\times 10^{-3}{.} 
\]

In this study of $B\rightarrow \rho l\nu $ decay, there is only one free
parameter i.e. $g_{+}\left( 0\right) $, and we have used the value of this
predicted by a LCSR calculation \cite{PRD58-094016} given in Eq. (\ref
{aaaaa1}). The error appear in our calculation from this parameter. The
error shown in our result of $\Gamma \left( B\rightarrow \rho l\nu \right) $
in (\ref{width-our}) is only 13.8\%.

\section{Numerical Analyses and Discussions \label{numerical-analyses}}

The values of the helicity amplitudes and form factors at $q^2=0$ and $%
q^2=q_{\max }^2$ are given in Tables \ref{hmohp}, \ref{qsqmax}, and \ref
{fftable}. The variation of form factors at $q^2=0$ and $q^2=q_{\max }^2$ is
shown in Table \ref{fftable}. None of the helicity amplitudes changes sign
over the entire physical range of $q^2$ as shown in Fig. \ref{fig-helicity}.
Although $H_{+}$ and $H_0$ both get contributions from two form factors
which enters with opposite signs. The sqaure of the helicity amplitudes is
plotted in Fig. \ref{fig-helicitysq}.

First, $H_{-}$ is generally dominent over $H_{+}$, while the condition $%
H_{-}=H_{+}$ is forced by kinematics at $q^2=q_{\max }^2$; these two
amplitudes go their own way and differ by more than a factor of 12.7 at $%
q^2=0$, a comparison over this point is given in Table \ref{hmohp}. 
\begin{table}[tbp]
\caption{A comparison between different calculations is given between the
helicity amplitudes $H_{-}$, $H_{+}$ (GeV) and their ratio $H_{-}/H_{+}$ at $%
q^2=0$.}
\label{hmohp}
\begin{center}
\begin{tabular}{|c|c|c|c|c|}
\hline
& $H_{+}\left( 0\right) $ & $H_{-}\left( 0\right) $ & $H_{-}\left( 0\right)
/H_{+}\left( 0\right) $ & $\sqrt{q^2}H_0\left( 0\right) $ \\ \hline
Our & 0.256 & 3.253 & 12.714 & 8.737 \\ \hline
QM \cite{PLB436-344} & 0.163 & 2.942 & 18.014 & 8.202 \\ \hline
LCSR \cite{PRD58-094016} & 0.055 & 3.103 & 56.767 & 10.158 \\ \hline
Lattice \cite{PLB416-392} & 0.0898 & 4.224 & 47.020 & 10.914 \\ \hline
\end{tabular}
\end{center}
\end{table}
\begin{table}[tbp]
\caption{A comparison between different calculations is given between the
helicity amplitudes $H_{-}$, $H_{+}$ (GeV) at $q^2=q_{\max }^2$.}
\label{qsqmax}
\begin{center}
\begin{tabular}{|c|c|c|c|c|}
\hline
& $H_{+}\left( q_{\max }^2\right) $ & $H_{-}\left( q_{\max }^2\right) $ & $%
H_0\left( q_{\max }^2\right) $ & $\sqrt{q^2}H_0\left( q_{\max }^2\right) $
\\ \hline
Our & 2.362 & 2.362 & 2.362 & 10.652 \\ \hline
QM \cite{PLB436-344} & 2.952 & 2.952 & 2.952 & 13.313 \\ \hline
LCSR \cite{PRD58-094016} & 2.782 & 2.782 & 2.782 & 12.549 \\ \hline
Lattice \cite{PLB416-392} & 2.826 & 2.826 & 2.826 & 12.744 \\ \hline
\end{tabular}
\end{center}
\end{table}
\begin{table}[tbp]
\caption{Variation of form factors at $q^2=0$ $\left( q^2=q_{\max }^2\right) 
$ is shown. A comparison of the ratio of respective form factors at $%
q^2=q_{\max }^2$ over at $q^2=0$ is presented.}
\label{fftable}
\begin{center}
\begin{tabular}{|c|c|c|c|c|c|c|}
\hline
& $V$ & $A_1$ & $A_2$ & $\frac{V\left( q_{\max }^2\right) }{V\left( 0\right) 
}$ & $\frac{A_1\left( q_{\max }^2\right) }{A_1\left( 0\right) }$ & $\frac{%
A_2\left( q_{\max }^2\right) }{A_2\left( 0\right) }$ \\ \hline
Our & 0.332 (2.605) & 0.29 (0.390) & 0.279 (1.136) & 7.840 & 1.346 & 4.06 \\ 
\hline
QM \cite{PLB436-344} & 0.308 (1.915) & 0.257 (0.488) & 0.242 (0.956) & 6.218
& 1.901 & 3.955 \\ \hline
LCSR \cite{PRD58-094016} & 0.338 (2.011) & 0.261 (0.460) & 0.223 (0.818) & 
5.948 & 3.135 & 2.062 \\ \hline
Lattice \cite{PLB416-392} & 0.458 (2.102) & 0.356 (0.467) & 0.342 (0.898) & 
4.587 & 1.310 & 2.628 \\ \hline
\end{tabular}
\end{center}
\end{table}
Again kinematics forces $H_0=H_{-}=H_{+}$ at $q^2=q_{\max }^2$, but the
amplitudes soon separate so that in the region near $q^2=0$ the amplitude $%
H_0$, which contain a factor of $1/\sqrt{q^2}$, very much dominates the
others. This feature is independent of modest changes in the form factors.

We have used vector--current form factor pole at 5.33 GeV and axial--vector
current form factor pole at 5.71 GeV. Note that the poles are very close to
the edge of physical region. This means that on the one hand, they should
indeed dominate the behavior of the form factors near $q^2=q_{\max }^2$. The
change in the $q^2$ range in different form factors give quite distinct
results as shown in Table \ref{fftable}.

The $H_{+}$ $\left( H_{-}\right) $ helicity amplitudes gets destructive
(constructive) contributions from the $A_1$ and $V$ form factors. The near
zero behavior in $H_{+}$ arises, since $V$ with its coefficient $2M_B\left| 
{\bf k}_\rho \right| /\left( M_B+M_\rho \right) $ in Eq. (\ref{cHpm}) causes
its contribution to this halicity amplitude to quickly catch up to that of $%
A_1$ as we move away from $q^2=q_{\max }^2$. Of course this increases $H_{-}$
correspondingly, according to Eq. (\ref{cHpm}).

A similar situation pertains to the helicity amplitude $H_0$, where $A_1$
and $A_2$ destructively interfere, as shown in Eq. (\ref{cH0}). This pulls
the whole longitudnal portion of the decay rate $\left( \Gamma _L=\Gamma
_0\right) $ down and it is far smaller than the transverse one $\left(
\Gamma _T=\Gamma _{+}+\Gamma _{-}\right) $, see Table \ref{decay-widths}.
This agrees with the observations of Gilman and Singleton \cite{PRD41-142},
and K\"{o}rner and Schuler \cite{ZPC38-511} (as shown in Eq. (19) of Ref. 
\cite{ZPC38-511}) but contradicts the recent observation of Ali and Safir 
\cite{0205254v3} in the LEET approach. 
\begin{table}[tbp]
\caption{Different decay rate values are calculated and compared with
different approaches. The comparison of $\Gamma _L/\Gamma _T$ is presented
and compare with other approaches.}
\label{decay-widths}
\begin{center}
\begin{tabular}{|c|c|c|c|c|c|c|}
\hline
& $\Gamma _{-}$ & $\Gamma _{+}$ & $\Gamma _L=\Gamma _0$ & $\Gamma _T$ & $%
\Gamma _{{Total}}$ & $\Gamma _L/\Gamma _T$ \\ \hline
Our & 9.662 & 0.059 & 6.095 & 9.721 & 15.82 & 0.63 \\ \hline
QM \cite{PLB436-344} & - & - & - & - & 15.8$\pm $2.3 & 0.88$\pm 0.08$ \\ 
\hline
LCSR \cite{PRD55-5561} & - & - & - & - & 13.5$\pm 4.0$ & 0.52$\pm $0.08 \\ 
\hline
Lattice \cite{PLB416-392} & - & - & - & - & 16.5$_{-2.3}^{+3.5}$ & 0.80$%
_{-0.03}^{+0.04}$ \\ \hline
ISGW2 \cite{ISGW2} & - & - & - & - & 14.2 & 0.3 \\ \hline
\end{tabular}
\end{center}
\end{table}
\begin{table}[tbp]
\caption{The values of the $V/A_1$ at different $q^2$ is given and compared
with HQET--LEET approach and with recent Lattice data. Also ratio $\Gamma
_{+}/\Gamma _{-}$ is given.}
\label{VoA1}
\begin{center}
\begin{tabular}{|c|c|c|c|c|}
\hline
Observable & $
\begin{array}{c}
V/A_1 \\ 
q^2=0
\end{array}
$ & $
\begin{array}{c}
V/A_1 \\ 
q^2=14.53\, { GeV}^2
\end{array}
$ & $
\begin{array}{c}
V/A_1 \\ 
q^2=14.68\, { GeV}^2
\end{array}
$ & $\Gamma _{+}/\Gamma _{-}$ \\ \hline
Our & 1.146 & 2.96 & 3.00 & 0.006 \\ \hline
HQET--LEET \cite{PRD60-014001} & 1.29 & 2.63 & 2.66 & 0.02 \\ \hline
Lattice \cite{lat0209116} & - & 2.05 & 2.02 & - \\ \hline
\end{tabular}
\end{center}
\end{table}

As an exercise, we check the relation among the form factors which holds
true in large energy limit (LEL) (for a recent discussion see Refs. \cite
{PRD60-014001,PRD63-113008,PRD63-114020,0206152}), namely 
\begin{equation}
\frac{A_1\left( q^2\right) }{V\left( q^2\right) }=\frac{2E_VM_B}{\left(
M_B+M_V\right) ^2}=\frac{M_B^2+M_V^2-q^2}{\left( M_B+M_V\right) ^2}{.}
\label{A1oV}
\end{equation}
In Fig. \ref{a1overv}, we plot the ratio of $B\rightarrow \rho $ form
factors $A_1\left( q^2\right) /V\left( q^2\right) $ and compare them with
the LEL \cite{PRD60-014001} Eq. (\ref{A1oV}) (dashed line), with the result
of Ref. \cite{9810421,PRD64-094022} (dotted line) and the lattice results 
\cite{lat0209116}, whereas the solid line is the result of this calculation.
The knowledge of this ratio has important consequences: on the one hand, its
measurment at $q^2=0$ will provide a test of various approaches; on the
other hand it provides the ratio $\Gamma _{+}/\Gamma _{-}$ of the width to
helicity eigenstates $\lambda =\pm 1$. In Table \ref{VoA1}, the values of
the $V/A_1$ at different $q^2$ is given and compared with HQET--LEET \cite
{PRD60-014001} approach and with recent Lattice data \cite{lat0209116}. Also
ratio $\Gamma _{+}/\Gamma _{-}$ is given in Table \ref{VoA1}.

\section{Conclusions\label{conclusion}}

We have studied $B\rightarrow \rho l\nu _l$ decay by using Ward Identities.
The form factors have been calculated and found that their normalization is
essentially determined by a single constant $g_{+}\left( 0\right) $. Then we
use a parametrazation [Eq. (\ref{cFqsq})] which take into account potential
corrections to single pole dominance arising from radial excitations of $M$
(where $M=M_{B^{*}}$ or $M_{B_A^{*}}$), as suggested by dispersion relations 
\cite{PLB214-459,ZPC41-217}. This also takes care of off mass shell--ness of
couplings $B^{*}$ or $B_A^{*}$ with $B\rho $ channel \cite{ZPC48-55} as
indicated in Eq. (\ref{replace}). We took the value of $g_{+}\left( 0\right)
=0.29\pm 0.04$ from a recent LCSR calculations \cite{PRD58-094016}. And
predict the value of $g_{B^{*}B\rho }=2\left( 1.26\pm 0.17\right) /f_B$ GeV$%
^{-1}$ [cf. Eq. (\ref{gBsBrho})], which agrees very well with CQM
calculation (\ref{gBsBrhoCQM}) and the value obtained by QCD sum rules $%
\left[ \sqrt{2}\left( 11\right) \,{ GeV}^{-1}\right] $. Also, we have
predicted a relation between $S$-wave and $D$-wave couplings $%
g_{B_A^{*}B\rho ^{-}}=2.66\times f_{B_A^{*}B\rho ^{-}}$ GeV$^2$ given in Eq.
(\ref{swave-Dwave}). The estimate for $A_0\left( 0\right) \simeq 0.320\pm
0.044$ is consistant with the value given in Eq. (\ref{A00a}) and with the
LCSR value 0.372 \cite{PRD58-094016}. We summarize the form factors in Eq. (%
\ref{ffvalues}) and their values at $q^2=0$ in Eq. (\ref{ff0values}).

A detailed analysis of form factors and helicity amplitudes is presented.
Total decay rate is obtained which is $15.82$ $\left| V_{ub}\right| ^2$ ps$%
^{-1}$ [cf. Eq. \ref{width-our}], which results in predicting the CKM matrix
element $\left| V_{ub}\right| $ 
\[
\left| V_{ub}\right| =3.24\times 10^{-3} 
\]
by using CLEO branching ratio result \cite{PRD61-052001}, and 
\[
\left| V_{ub}\right| =3.67\times 10^{-3} 
\]
by using BaBar branching ratio result \cite{ex0301001}. While using the
branching ratio of recent study of CLEO \cite{ex0304019}, we obtain the
value of $\left| V_{ub}\right| $ 
\[
\left| V_{ub}\right| =2.98\times 10^{-3}{.} 
\]
Our results of $\left| V_{ub}\right| $ agree very well with the respective
results of $\left| V_{ub}\right| $ of CLEO \cite{PRD61-052001}, and the
recent results of BaBar \cite{ex0301001}, and CLEO \cite{ex0304019}. The
error shown in our result of $\Gamma \left( B\rightarrow \rho l\nu \right) $
in (\ref{width-our}) is only 13.8\%.

\appendix 

\section{Derivation of Ward Identities}

We start with the relation 
\begin{equation}
\left\langle V\left( k,\epsilon \right) \left| i\bar{u}\sigma ^{\mu \nu
}q_\nu \gamma _5b\right| B\left( p\right) \right\rangle e^{-i\left(
p-k\right) \cdot x}=-\left\langle V\left( k,\epsilon \right) \left| \partial
_\nu \left( \bar{u}\left( x\right) \sigma ^{\mu \nu }\gamma _5b\left(
x\right) \right) \right| B\left( p\right) \right\rangle  \label{A1}
\end{equation}
We can replace $\partial _\nu $ by the covariant derivative $D_\nu $ to take
into account the strong interaction of the the quark field. We can write the
right hand side of Eq. (\ref{A1}) as 
\begin{equation}
-\left\langle V\left( k,\epsilon \right) \left| D_\nu \bar{u}\left( x\right)
\left[ -i\gamma ^\nu \gamma ^\mu +ig^{\mu \nu }\right] \gamma _5b\left(
x\right) \right| B\left( p\right) \right\rangle -\left\langle V\left(
k,\epsilon \right) \left| \bar{u}\left( x\right) \left[ i\gamma ^\mu \gamma
^\nu -ig^{\mu \nu }\right] \gamma _5D_\nu b\left( x\right) \right| B\left(
p\right) \right\rangle  \label{A2}
\end{equation}
Using Dirac equation 
\begin{equation}
\not{D}b\left( x\right) =-im_bb\left( x\right) ,\,\,\,\,\,\,\,\bar{u}\left(
x\right) \not{D}=im_u\bar{u}\left( x\right)  \label{A3}
\end{equation}
the relation (\ref{A2}) becomes 
\begin{eqnarray}
&&\left( m_b-m_u\right) \left\langle V\left( k,\epsilon \right) \left| \bar{u%
}\left( x\right) \gamma ^\mu \gamma _5b\left( x\right) \right| B\left(
p\right) \right\rangle  \nonumber \\
&&-i\left\langle V\left( k,\epsilon \right) \left| D^\mu \bar{u}\left(
x\right) \gamma _5b\left( x\right) \right| B\left( p\right) \right\rangle 
\nonumber \\
&&+i\left\langle V\left( k,\epsilon \right) \left| \bar{u}\left( x\right)
\gamma _5D^\mu b\left( x\right) \right| B\left( p\right) \right\rangle 
\nonumber \\
&=&\left( m_b-m_u\right) \left\langle V\left( k,\epsilon \right) \left| \bar{%
u}\gamma ^\mu \gamma _5b\right| B\left( p\right) \right\rangle e^{-iq\cdot x}
\nonumber \\
&&-i\left\langle V\left( k,\epsilon \right) \left| D^\mu \left( \bar{u}%
\left( x\right) \gamma _5b\left( x\right) \right) \right| B\left( p\right)
\right\rangle  \nonumber \\
&&+2i\left\langle V\left( k,\epsilon \right) \left| \bar{u}\left( x\right)
\gamma _5D^\mu b\left( x\right) \right| B\left( p\right) \right\rangle
\label{A4}
\end{eqnarray}
Using now the linear momentum commutation relation 
\[
\left[ \hat{P}^\mu ,O\left( x\right) \right] =-iD^\mu O\left( x\right)
,\,\,\,\,\hat{P}_q^\mu =\int d^3x:q^{\dagger }\left( x\right) D^\mu q\left(
x\right) : 
\]
and 
\begin{eqnarray}
\left\langle V\left( k,\epsilon \right) \right| \hat{P}^\mu &=&k^\mu
\left\langle V\left( k,\epsilon \right) \right|  \nonumber \\
\hat{P}^\mu \left| B\left( p\right) \right\rangle &=&p^\mu \left| B\left(
p\right) \right\rangle ,  \label{A5}
\end{eqnarray}
the last two terms in Eq. (\ref{A4}) become 
\begin{equation}
-q^\mu \left\langle V\left( k,\epsilon \right) \left| \bar{u}\gamma
_5b\right| B\left( p\right) \right\rangle e^{-iq\cdot x}+2\left\langle
V\left( k,\epsilon \right) \left| \bar{u}\gamma _5bp_b^\mu \right| B\left(
p\right) \right\rangle e^{-iq\cdot x}  \label{A6}
\end{equation}
where in the last term we have used that $P_b^\mu \left| V\left( k,\epsilon
\right) \right\rangle =0$ as $V\left( k,\epsilon \right) $ does not contain
the quark $b$. In the heavy quark effective theory $m_b$ is taken to
infinity in a way that $p_b^\mu /m$ is held fixed, but the four momentum of
the light degree of freedom are neglected compared with $m_b$ \cite
{PRD42-2388}. This enables us to identify $p_b^\mu $ with $p^\mu $. Hence
from Eqs. (\ref{A1}), (\ref{A4}) and (\ref{A6}), we obtain the Ward identity
[Eq. (\ref{WI1})] $\left[ 2p-q=p\rightarrow k\right] $%
\begin{eqnarray}
\left\langle V\left( k,\epsilon \right) \left| i\bar{u}\sigma ^{\mu \nu
}q_\nu \gamma _5b\right| B\left( p\right) \right\rangle &=&\left(
m_b-m_u\right) \left\langle V\left( k,\epsilon \right) \left| \bar{u}\gamma
^\mu \gamma _5b\right| B\left( p\right) \right\rangle  \nonumber \\
&&+\left( p^\mu +k^\mu \right) \left\langle V\left( k,\epsilon \right)
\left| \bar{u}\gamma _5b\right| B\left( p\right) \right\rangle  \nonumber \\
&&  \label{A8}
\end{eqnarray}
Similarly, 
\begin{eqnarray}
\left\langle V\left( k,\epsilon \right) \left| i\bar{u}\sigma ^{\mu \nu
}q_\nu b\right| B\left( p\right) \right\rangle &=&-\left( m_b-m_u\right)
\left\langle V\left( k,\epsilon \right) \left| \bar{u}\gamma ^\mu b\right|
B\left( p\right) \right\rangle  \nonumber \\
&&+\left( p^\mu +k^\mu \right) \left\langle V\left( k,\epsilon \right)
\left| \bar{u}b\right| B\left( p\right) \right\rangle  \nonumber \\
&&  \label{A9}
\end{eqnarray}
However, the matrix element $\left\langle V\left( k,\epsilon \right) \left| 
\bar{u}b\right| B\left( p\right) \right\rangle $ vanishes due to parity
consideration. Thus, we obtain the Ward Identity (\ref{WI1a}) given in the
text.

\underline{{\bf Acknowledgements}}: Two of us (R and A) would like to
acknowledge the support of King Fahd University of Petroleum and Minerals.
The work of G and R was supported by Pakistan Council for Science and
Technology. One of us (G) like to thank Professor Fayyazuddin for many
helpful discussions over the subject. (G) also thanks M.H. Nasim, Zahoor
Ahmad, and M. Shafiq for their ever ready help at different stages of work.

\section{Figure Captions}

\begin{enumerate}
\item  Comparison of the form factor vs. lattice data \cite{BaBar,lat9611016}
and calculations within different approaches is given. Solid line--our
result, dotted line--LCSR \cite{PRD58-094016}, dashed line--quark model \cite
{PLB436-344}, dash-dotted line--lattice constrained parametriation of \cite
{PLB416-392}.\label{figva1c}

\item  Comparison of Helicity amplitudes vs calculations within different
approaches is given. Line description is as in Fig. \ref{figva1c}.\label
{fig-helicity}

\item  $d\Gamma /dq^2$ distribution for each of the three terms in Eq. (\ref
{ddhelicitysq}): (a) the terms proportional to $\left| H_{-}\right| ^2$ and $%
\left| H_{+}\right| ^2$, (b) the terms proportional to $\left| H_0\right| ^2$%
. Line description is as in Fig. \ref{figva1c}. \label{fig-helicitysq}

\item  Ratio of $B\rightarrow \rho $ form factors, $A_1\left( q^2\right)
/V\left( q^2\right) $, are plotted with lattice data points \cite{lat0209116}%
. The solid line is our result. The dashed line is the LEET prediction \cite
{PRD60-014001} and the dotted line is the prediction of \cite
{9810421,PRD64-094022}.\label{a1overv}
\end{enumerate}


\begin{thebibliography}{99}
\bibitem{dgplus0}  With $\xi _{\parallel }=\xi _{\perp }$, Soares \cite
{9810421} finds a good agreement between his prediction and the lattice data 
\cite{NPB447-425}. In the Bakamjian-Thomas formalism \cite{PR92-1300}, $\xi
_{\parallel }=\xi _{\perp }$ would hold if the $D$-wave were asymptotically
suppressed by some power of $E$ with respect to the $S$-wave in the vector
meson \cite{PLB451-187}.
\end{thebibliography}
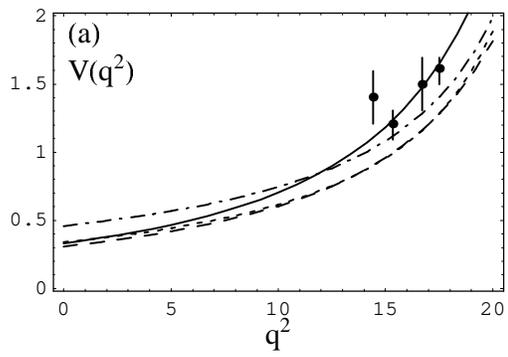
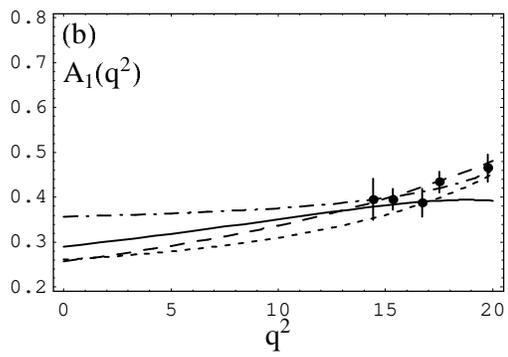
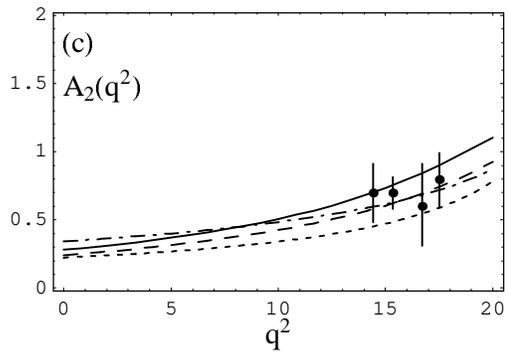
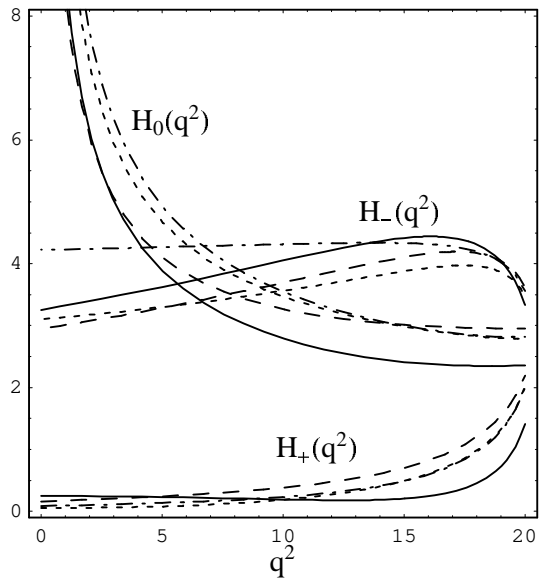
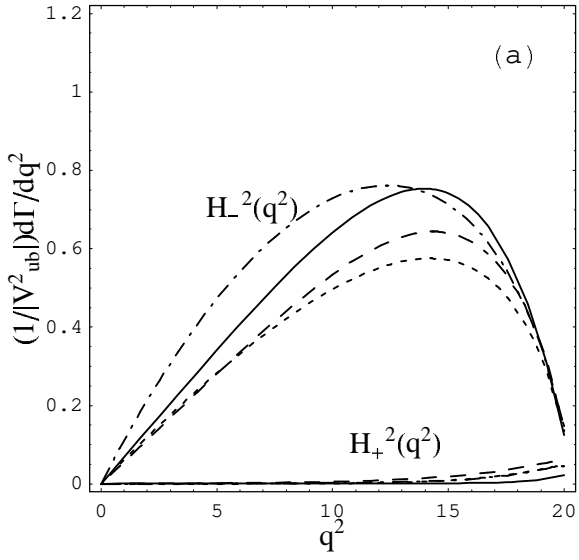
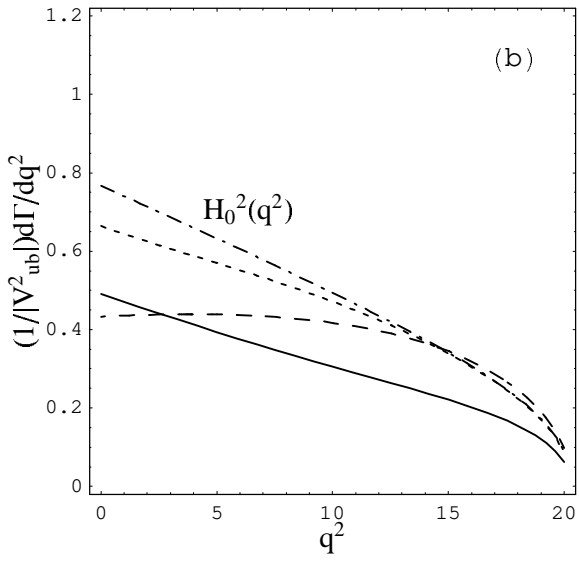
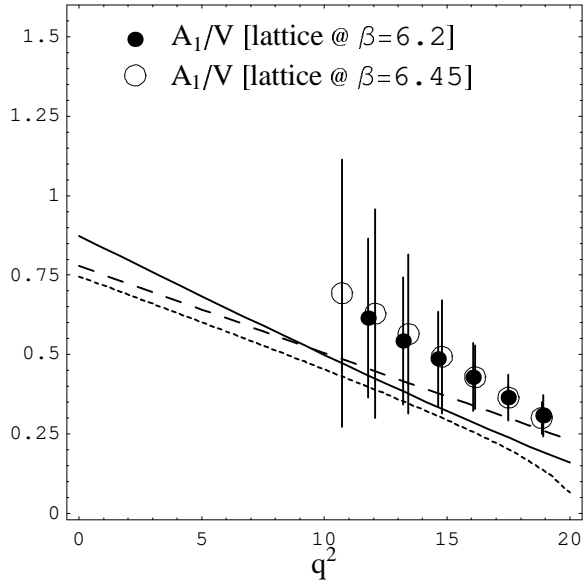
\end{document}